\documentclass{article}

\usepackage[final, main]{neurips_2025}

\usepackage[utf8]{inputenc}
\usepackage[T1]{fontenc}
\usepackage{hyperref}
\usepackage{url}
\usepackage{booktabs}
\usepackage{amsmath, amsfonts}
\usepackage{microtype}
\usepackage{xcolor}
\usepackage{enumitem}
\usepackage{graphicx}
\setlist[itemize]{leftmargin=10pt,topsep=0pt}
\setlist[enumerate]{leftmargin=10pt}

\makeatletter
\renewcommand\paragraph{\@startsection{paragraph}{4}{\z@}%
  {\parskip}% space before (same as paragraph spacing)
  {-1em}% space after (run-in)
  {\normalfont\normalsize\bfseries}}
\makeatother

\title{ClusterFusion++: Expanding Cluster-Level Fusion to Full Transformer-Block Decoding}

\author{%
  Chiheng Jin \quad Hongche Yu \quad Xihui Chen\\
  Shanghai Jiao Tong University\\
  \texttt{\{wendy-hamlet, superk977, charly-a\}@sjtu.edu.cn}\\
}

\begin{document}
\maketitle

\begin{abstract}
Large language model (LLM) decoding is latency-sensitive and often bottlenecked by fragmented operator execution and repeated off-chip materialization of intermediate tensors. Prior work \citep{luo2025clusterfusion} expands fusion scope by leveraging thread-block clusters and on-chip inter-block collectives to fuse attention-side operators (QKV projection, attention, and output projection). We develop \textbf{ClusterFusion++}, a \textbf{CUDA-level} extension that broadens fusion to the \textbf{full Transformer decoder block} for GPT-NeoX/Pythia models: \emph{LayerNorm $\rightarrow$ QKV $\rightarrow$ RoPE $\rightarrow$ decode attention $\rightarrow$ output projection $\rightarrow$ Post-LN $\rightarrow$ MLP $\rightarrow$ residual}. We additionally engineer a CUDA-Graph-compatible execution mode with persistent Tensor Memory Accelerator (TMA) descriptors to reduce per-step overhead. On an NVIDIA RTX 5090-class GPU, ClusterFusion++ improves throughput by 1.34\(\times\) for Pythia-2.8B and yields similar gains for Pythia-6.9B, while maintaining high output fidelity (near-token-identical generation, with minor non-determinism from FP16 atomics).
Our code is open-sourced at \url{https://github.com/superk668/ClusterFusionPlus}.
\end{abstract}

\section{Introduction}
Large language models (LLMs) have become the backbone of modern artificial intelligence, whose applications span across various domains \citep{liu2024exploring} \citep{yang2024qwen} \citep{su2022contrastive}. Autoregressive decoding dominates end-to-end latency for many LLM workloads as context length and model size grow. Its acceleration has been a major focus of the research community, with many techniques being proposed to improve the efficiency of LLM inference \citep{dao2023flashdecoding} \citep{yang2024pyramidinfer} \citep{ainslie2023gqa} \citep{wu2024layer} \citep{nvidia2024tensorrtllm}.

In common inference stacks, a single decoding step of one Transformer block decomposes into many GPU kernels, each producing intermediate tensors that are written to and read from global memory. This fragmented execution incurs heavy off-chip traffic and kernel-launch overhead, both of which limit practical latency.

While the optimization of this execution dataflow has been widely studied \citep{hong2024flashdecoding} \citep{liu2025flashmla}, the introduction of recent GPU architecture features has enabled new opportunities for optimization. \textbf{ClusterFusion} \citep{luo2025clusterfusion} observes that modern NVIDIA GPUs expose \emph{thread-block clusters} with \emph{distributed shared memory} (DSMEM), enabling low-latency communication between blocks within a cluster \citep{nvidia2024hopper}. By providing cluster-level collective primitives, ClusterFusion fuses attention-side operators into a single cluster-coordinated kernel, reducing off-chip intermediate traffic.

In this project we ask: \emph{can cluster-enabled fusion be extended from attention-side fusion to the full Transformer decoder block in a real GPT-style model?} We answer yes, and in our project \textbf{ClusterFusion++}, we make the following contributions:
\begin{itemize}
  \item \textbf{Full-block fusion for GPT-NeoX/Pythia}: We port ClusterFusion-style cluster-centric decoding to the Pythia family (GPT-NeoX architecture) \citep{biderman2023pythia} and expand fusion scope including LayerNorm, attention, partial RoPE, Post-LN and MLP, which yields a full decoder-block fused kernel for decoding.
  \item \textbf{Architecture-aware kernel mapping}: We handle Pythia-2.8B's non-power-of-two head dimension (\(d_\text{head}=80\)) with warp/tiling choices that preserve correctness and performance.
  \item \textbf{CUDA Graph execution mode}: We implement a reusable graph context that creates TensorMap (TMA) descriptors once per layer and reuses static buffers across decode steps, reducing per-step setup overhead.
\end{itemize}

\section{Methods}
\subsection{Background: decoding operators and fusion}
A Transformer decoder block for a single token typically performs:
\emph{LayerNorm $\rightarrow$ QKV $\rightarrow$ RoPE $\rightarrow$ decode attention $\rightarrow$ output projection $\rightarrow$ MLP with residual}. In standard execution, each step is scheduled as a separate kernel (or a few kernels), forcing intermediate tensors through global memory. \textbf{Fusion} means executing multiple consecutive operators inside one kernel, keeping intermediate values in registers/shared memory instead of materializing them in global memory. However, fusion is traditionally limited by inter-block dependencies: when a result requires a reduction across blocks, frameworks typically end a kernel and use global memory as the rendezvous point.

With recent GPU architecture features \citep{nvidia2024hopper}, we can fuse more operators inside a single kernel. Thread-block clusters allow a set of blocks to be co-scheduled with fast inter-block communication through DSMEM. ClusterFusion++ follows the cluster-centric philosophy of ClusterFusion \citep{luo2025clusterfusion}: blocks within a cluster collaboratively compute and exchange partial results on-chip, enabling larger fused regions than block-isolated kernels.

\subsection{ClusterFusion++: Full-block fusion for GPT-NeoX architecture}

Building upon ClusterFusion's attention-side fusion, ClusterFusion++ extends the fused region to cover the \emph{entire} decoder block for GPT-NeoX/Pythia architectures. A single kernel invocation performs:
\emph{Pre-attention LayerNorm $\rightarrow$ QKV projection and KV cache update $\rightarrow$ Rotary position embedding (RoPE) $\rightarrow$ Decode attention over KV cache $\rightarrow$ Output projection with residual connection $\rightarrow$ Post-attention LayerNorm $\rightarrow$ MLP (up-projection, GELU, down-projection) with residual connection}.

\paragraph{Architecture-Specific Adaptations}
ClusterFusion++ introduces several architecture-specific adaptations to support the GPT-NeoX/Pythia architecture. We add bias terms to the LayerNorm, QKV projection, and MLP layers, and the weights are stored interleaved per head rather than concatenated. ClusterFusion++ also supports the unique RoPE in Pythia-2.8B, where only the first 25\% of each head dimension undergoes rotation, and the remaining dimensions are passed through unchanged.

\paragraph{Kernel Optimizations}
Beyond adaptation, we introduce the following kernel optimizations:
\begin{itemize}
  \item \textbf{Single-pass LayerNorm}: We compute mean and variance simultaneously using $\mathrm{Var}(x)=\mathbb{E}[x^2]-\mathbb{E}[x]^2$, halving memory traffic compared to two-pass implementations.
  \item \textbf{Cluster-cooperative attention}: Multiple thread blocks within a cluster cooperatively cover the KV cache sequence length, with efficient cross-block communication via distributed shared memory.
  \item \textbf{Tree reduction for output accumulation}: We optimize cluster-level reduction from sequential ring reduction ($O(n)$ steps) to tree reduction ($O(\log n)$ steps), reducing synchronization overhead by parallelizing the reduction through a binary tree structure.
  \item \textbf{PTX-accelerated GELU}: We use inline PTX intrinsics to compute GELU activation with reduced instruction count.
\end{itemize}

\subsection{CUDA Graph Mode with Persistent TensorMaps}

To minimize per-token overhead during autoregressive decoding, we implement a CUDA Graph context for each layer.
It creates TensorMap (TMA) descriptors once per layer and reuses static buffers across decode steps, reducing per-step setup overhead.
Buffers, including output and intermediate tensors, are allocated once and reused across decode iterations.
The decode step then becomes a single graph replay, eliminating CPU-side kernel-launch overhead.
This complements operator fusion by reducing both GPU kernel overhead and CPU dispatch latency.

\section{Experiments}
\paragraph{Experimental Setup}
We evaluate ClusterFusion++ on an NVIDIA RTX 5090-class GPU (\texttt{sm\_120}) using Pythia-2.8B and Pythia-6.9B models (GPT-NeoX). Sequence length ranges from 16 to 2048, and batch size is 1. All experiments use PyTorch 2.9.1 and CUDA 13.1. Our baseline is HuggingFace Transformers decoding with KV cache enabled.
We evaluate ClusterFusion++ on two models: Pythia-2.8B and Pythia-6.9B, both based on the GPT-NeoX architecture.

\begin{figure}[h]
\centering
\begin{minipage}{0.62\textwidth}
\centering
\includegraphics[width=\textwidth]{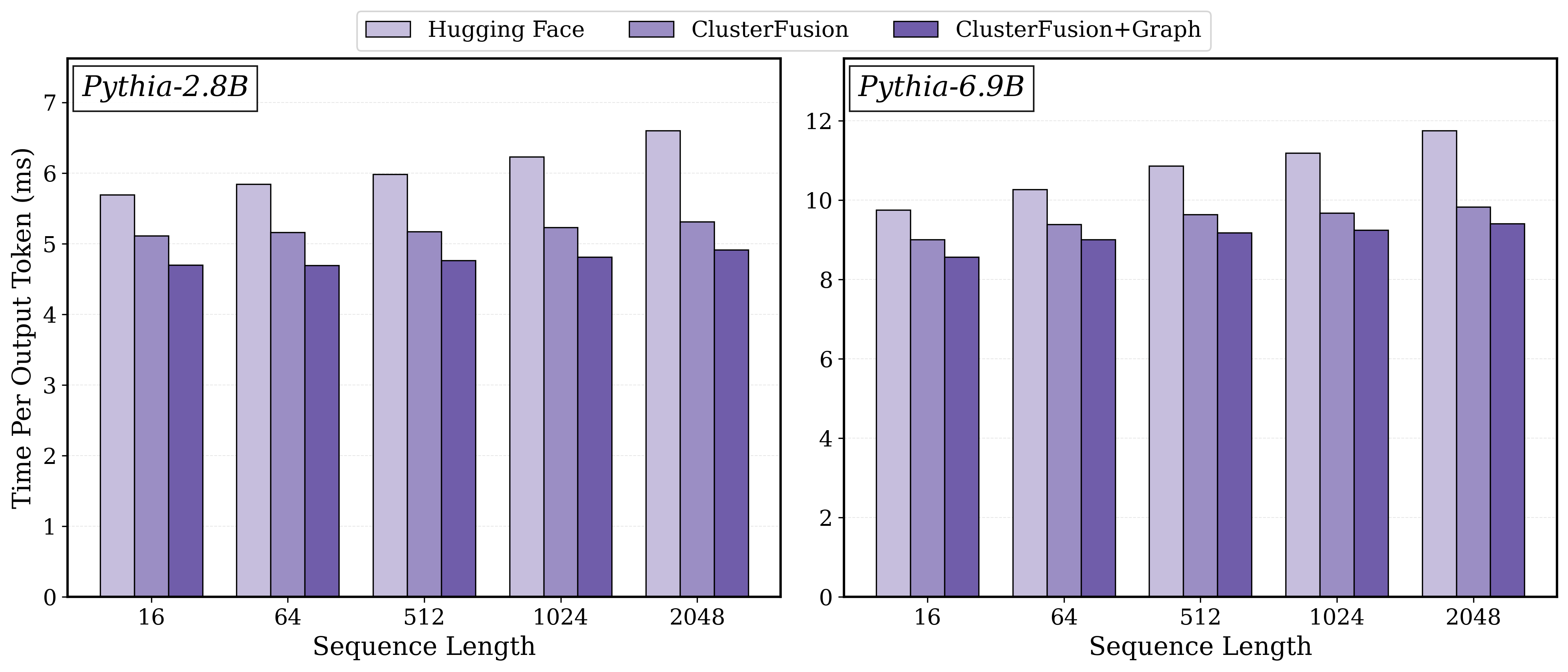}
\caption{Time per output token (TPOT) of Pythia-2.8B (left) and Pythia-6.9B (right) on RTX 5090.}
\label{fig:tpot}
\end{minipage}
\hfill
\begin{minipage}{0.35\textwidth}
\centering
\includegraphics[width=\textwidth]{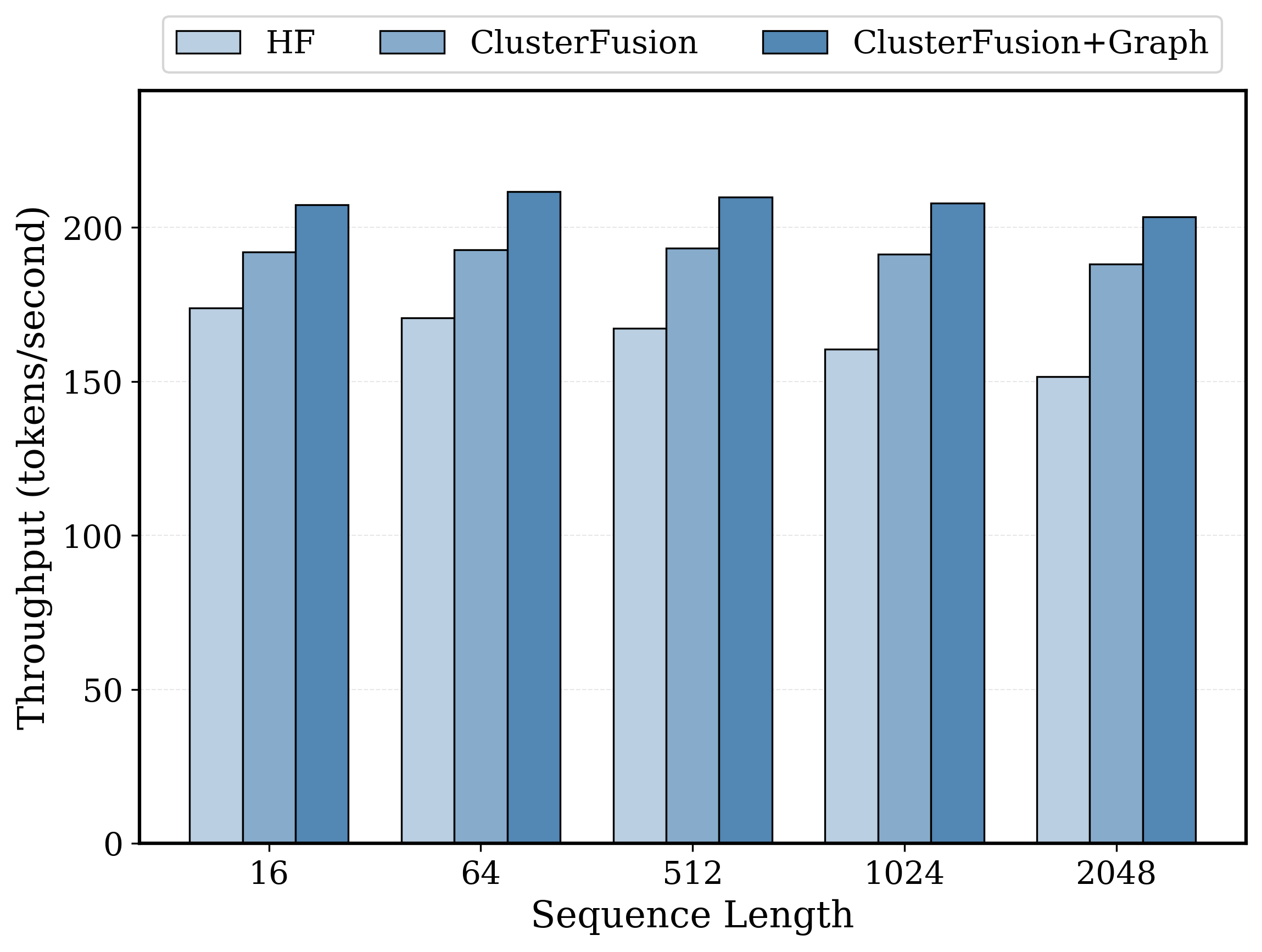}
\caption{Throughput of Pythia-2.8B on RTX 5090.}
\label{fig:throughput}
\end{minipage}
\end{figure}

\paragraph{Results}
We use time per output token (TPOT) and throughput as the metrics for end-to-end evaluation.
Results appear in Figure \ref{fig:tpot} and Figure \ref{fig:throughput}.
For TPOT, ClusterFusion++ achieves 1.21$\times$, 1.25$\times$, 1.26$\times$, 1.30$\times$, and 1.34$\times$ speedup at different sequence lengths over the baseline for Pythia-2.8B, and 1.19$\times$, 1.24$\times$, 1.26$\times$, 1.29$\times$, and 1.34$\times$ for Pythia-6.9B.
See the Appendix for details.

\paragraph{Output fidelity.}
As shown in Table ~\ref{tab:ppl} and ~\ref{tab:qual_eval}, we observe near-token-identical generation for  prompts, with occasional mismatches attributable to FP16 atomic accumulation in the output projection. This behavior aligns with known non-determinism in parallel reductions with floating-point atomics.

\begin{table}[h]
\centering

\begin{minipage}{0.48\textwidth}
\centering
\vspace{2pt}
\caption{PPL on WikiText-2 and PG-19 (unchange from baseline because our kernel changes only accelerate the decode phase).}
\begin{tabular}{lrrrl}
\toprule
Dataset & PPL & Samples & Tokens \\
\midrule
WikiText-2 & 24.00 & 100 & 16,807  \\
PG-19 & 8.90 & 1 & 23,284\\
\bottomrule
\end{tabular}
\label{tab:ppl}
\end{minipage}
\hfill
\begin{minipage}{0.48\textwidth}
\centering
\vspace{2pt}
\caption{Quality evaluation on WikiText-2.}
\begin{tabular}{lccp{0.38\textwidth}}
\toprule
Metric & Overall & WikiText-2 \\
\midrule
Token Match Rate & 99.4\% & 99.8\% \\
Logits MAE & 0.0235 & -- \\
Top-5 Agreement & 92.3\% & 96.0\%  \\
Top-10 Agreement & 92.3\% & --  \\
\bottomrule
\end{tabular}
\label{tab:qual_eval}
\end{minipage}
\label{tab:quality}
\end{table}

\section{Ablation Study and Discussion}
\subsection{Decode Phase}
The decoding kernel is split into two components owned by two contributors: one handles the attention portion and the other the MLP. The components are later concatenated into a complete decoding kernel. We evaluate the performance improvement from each component separately, and Table ~\ref{tab:ablation} shows an interesting phenomenon: while the MLP down-projection kernel alone is \textbf{slower} than PyTorch's cuBLAS implementation (0.75$\times$ average speedup), combining it with the Attention+MLP-Up kernel yields better end-to-end performance than accelerating attention alone.

\begin{table}[h]
\centering
\caption{TPOT of different kernel configurations on Pythia-2.8B (with sequence length 2048).}
\begin{tabular}{lcc}
\toprule
\textbf{Configuration} & \textbf{Avg TPOT (ms)} & \textbf{vs PyTorch} \\
\midrule
PyTorch Baseline & 6.80 & 1.00$\times$ \\
CUDA Attention + PyTorch MLP Down & 5.32 & 1.28$\times$ \\
PyTorch Attention + CUDA MLP Down & 9.04 & 0.75$\times$ \\
Full Fused Kernel & 4.90 & 1.39$\times$ \\
\bottomrule
\end{tabular}

\label{tab:ablation}
\end{table}

\paragraph{Why MLP down alone is slower but provides synergy when fused.}
The standalone MLP Down kernel underperforms because of cuBLAS efficiency in PyTorch's \texttt{F.linear}, poorly amortized fixed overheads (TMA descriptor creation and cluster launch), and the memory-bound nature of loading 26.2M weight parameters.
However, when the MLP Down kernel is fused with the preceding Attention+MLP-Up operations, the combined kernel achieves 1.39$\times$ speedup---better than the 1.28$\times$ from attention-only acceleration. This synergy arises from amortized launch overhead, register/shared memory reuse of the 20KB MLP intermediate tensor, eliminated synchronization, and shared TMA infrastructure for weight loading. The memory-traffic reduction from fusion eliminates $2 \times 10240 \times 2 \times 32 = 1.31$ MB per decode step. At 1.8 TB/s memory bandwidth (RTX 5090), this saves approximately 0.73 ms, closely matching the observed improvement from 5.32 ms to 4.90 ms.

This demonstrates that kernel fusion benefits are \textbf{non-additive}: components individually slower than baseline can contribute positively when fused by eliminating intermediate memory traffic and amortizing fixed overheads.

\subsection{Prefill Phase}
For the prefill phase, we implement Flash Attention \citep{dao2022flashattention} and improve it on the GPT-NeoX architecture. However, while we do observe a speedup in Time To First Token (TTFT) of 1.56$\times$ over the PyTorch baseline for applying Flash Attention, our architecture-specific adaptation only achieve a speedup of 0.33$\times$. The latency is high due to pytorch-level limitations, but our variant still outperforms the baseline on memory efficiency and serves two important roles: (i) it isolates the memory benefit of the algorithm itself, proving that the idea works even in high-level frameworks, and (ii) it provides a transparent, hardware-agnostic reference that is easy to study, verify, and extend for future research.

\section{Conclusion}
ClusterFusion++ presents a CUDA-level cluster-centric fusion approach that expands ClusterFusion-style decoding fusion from attention-side operators to the full Transformer decoder block for GPT-NeoX/Pythia, enabling on-chip inter-block collectives via distributed shared memory to reduce intermediate global-memory traffic and launch overhead. Combined with a CUDA-Graph mode that reuses persistent TensorMap (TMA) descriptors and static buffers across decode steps, ClusterFusion++ outperforms the HuggingFace baseline on an RTX 5090 GPU across different configurations and models, while maintaining high output fidelity with only minor non-determinism attributable to FP16 atomic accumulation in cluster reductions.

\section*{Acknowledgments}
We thank the authors of ClusterFusion \citep{luo2025clusterfusion} for releasing their paper and codebase, which this project builds upon.

\appendix

\section{Team Work as a class project}
The ClusterFusion++ project \url{https://github.com/superk668/ClusterFusionPlus} is divided into three parts: prefill, decode-attention and decode-mlp.
Xihui Chen is in charge of the prefill phase \url{https://github.com/Sougetsusou/CS3602_project_ClusterFusion}, and Chiheng Jin is in charge of the decode-attention part \url{https://github.com/Wendy-Hamlet/CS3602_project_ClusterFusion} and Hongche Yu is in charge of the decode-mlp part \url{https://github.com/superk668/ClusterFusionPlus-MLP}.
The integration of the three parts is done by Chiheng Jin,
and this thesis is written by Hongche Yu.

\section{Detailed Data}

All benchmarks run on NVIDIA RTX 5090 (sm\_120), batch=1.

\begin{table}[h]
\centering
\caption{TPOT (Time Per Output Token) - Decode Phase for Pythia-2.8B}
\label{tab:tpot_decode}
\begin{tabular}{rrrrrr}
\toprule
Decode Tokens & HF (ms) & CF (ms) & CF+Graph (ms) & CF Speedup & Graph Speedup \\
\midrule
16 & 5.69 & 5.11 & 4.70 & 1.11$\times$ & \textbf{1.21$\times$} \\
32 & 5.70 & 5.11 & 4.69 & 1.12$\times$ & \textbf{1.22$\times$} \\
64 & 5.84 & 5.16 & 4.69 & 1.13$\times$ & \textbf{1.25$\times$} \\
128 & 5.76 & 5.11 & 4.69 & 1.13$\times$ & \textbf{1.23$\times$} \\
256 & 5.82 & 5.15 & 4.74 & 1.13$\times$ & \textbf{1.23$\times$} \\
512 & 5.98 & 5.17 & 4.76 & 1.16$\times$ & \textbf{1.26$\times$} \\
1024 & 6.23 & 5.23 & 4.81 & 1.19$\times$ & \textbf{1.30$\times$} \\
2048 & 6.60 & 5.31 & 4.91 & 1.24$\times$ & \textbf{1.34$\times$} \\
\bottomrule
\end{tabular}
\end{table}

\begin{table}[h]
\centering
\caption{Throughput (tokens/second) for Pythia-2.8B}
\label{tab:throughput_pythia28b}
\begin{tabular}{rrrrrr}
\toprule
Decode Tokens & HF & CF & CF+Graph & CF Speedup & Graph Speedup \\
\midrule
16 & 173.86 & 192.01 & 207.45 & 1.10$\times$ & \textbf{1.19$\times$} \\
32 & 174.39 & 193.78 & 210.54 & 1.11$\times$ & \textbf{1.21$\times$} \\
64 & 170.70 & 192.73 & 211.70 & 1.13$\times$ & \textbf{1.24$\times$} \\
128 & 173.37 & 195.16 & 212.36 & 1.13$\times$ & \textbf{1.22$\times$} \\
256 & 171.60 & 194.09 & 210.62 & 1.13$\times$ & \textbf{1.23$\times$} \\
512 & 167.21 & 193.36 & 209.90 & 1.16$\times$ & \textbf{1.26$\times$} \\
1024 & 160.51 & 191.29 & 207.85 & 1.19$\times$ & \textbf{1.29$\times$} \\
2048 & 151.50 & 188.21 & 203.46 & 1.24$\times$ & \textbf{1.34$\times$} \\
\bottomrule
\end{tabular}
\end{table}

\begin{table}[h]
\centering
\caption{FLOPs Estimation for Pythia-2.8B}
\label{tab:flops_pythia28b}
\begin{tabular}{rrrrr}
\toprule
Decode Tokens & Prefill (GFLOPs) & Decode (GFLOPs) & Total (GFLOPs) & TFLOPS/s (CF+Graph) \\
\midrule
16 & 26.48 & 84.78 & 111.26 & 1.44 \\
32 & 26.48 & 169.65 & 196.12 & 1.29 \\
64 & 26.48 & 339.63 & 366.11 & 1.21 \\
128 & 26.48 & 680.63 & 707.10 & 1.17 \\
256 & 26.48 & 1366.70 & 1393.18 & 1.15 \\
512 & 26.48 & 2755.22 & 2781.69 & 1.14 \\
1024 & 26.48 & 5597.68 & 5624.15 & 1.14 \\
2048 & 26.48 & 11544.32 & 11570.79 & 1.15 \\
\bottomrule
\end{tabular}
\end{table}

\begin{table}[h]
\centering
\caption{Pythia-6.9B Benchmark Results}
\label{tab:pythia69b_benchmark}
\begin{tabular}{rrrrrr}
\toprule
Decode Tokens & CF (s) & CF+Graph (s) & HF (s) & CF Speedup & Graph Speedup \\
\midrule
16 & 0.144 & 0.137 & 0.156 & 1.09$\times$ & \textbf{1.14$\times$} \\
32 & 0.296 & 0.284 & 0.323 & 1.09$\times$ & \textbf{1.14$\times$} \\
64 & 0.601 & 0.576 & 0.657 & 1.09$\times$ & \textbf{1.14$\times$} \\
128 & 1.224 & 1.160 & 1.338 & 1.09$\times$ & \textbf{1.15$\times$} \\
256 & 2.444 & 2.331 & 2.718 & 1.11$\times$ & \textbf{1.17$\times$} \\
512 & 4.935 & 4.699 & 5.562 & 1.13$\times$ & \textbf{1.18$\times$} \\
1024 & 9.910 & 9.464 & 11.453 & 1.16$\times$ & \textbf{1.21$\times$} \\
2048 & 20.135 & 19.256 & 24.065 & 1.20$\times$ & \textbf{1.25$\times$} \\
\bottomrule
\end{tabular}
\end{table}

\begin{table}[h]
\centering
\caption{End-to-End Benchmark for Attention-only Kernel (vs PyTorch Baseline)}
\label{tab:attention_only}
\begin{tabular}{rrrrrr}
\toprule
Tokens & CF(s) & PyTorch(s) & Speedup & TPOT CF(ms) & TPOT PT(ms) \\
\midrule
16 & 0.078 & 0.100 & 1.27$\times$ & 5.23 & 6.64 \\
32 & 0.163 & 0.209 & 1.28$\times$ & 5.25 & 6.74 \\
64 & 0.333 & 0.420 & 1.26$\times$ & 5.29 & 6.66 \\
128 & 0.668 & 0.845 & 1.26$\times$ & 5.26 & 6.65 \\
256 & 1.349 & 1.701 & 1.26$\times$ & 5.29 & 6.67 \\
512 & 2.724 & 3.463 & 1.27$\times$ & 5.33 & 6.78 \\
1024 & 5.518 & 7.125 & 1.29$\times$ & 5.39 & 6.96 \\
2048 & 11.249 & 14.950 & 1.33$\times$ & 5.50 & 7.30 \\
\bottomrule
\end{tabular}
\end{table}

\begin{table}[h]
\centering
\caption{End-to-End Benchmark for MLP-only Kernel (vs PyTorch Baseline)}
\label{tab:mlp_only}
\begin{tabular}{rrrrrr}
\toprule
Tokens & CF(s) & PyTorch(s) & Speedup & TPOT CF(ms) & TPOT PT(ms) \\
\midrule
16 & 0.134 & 0.098 & 0.73$\times$ & 8.96 & 6.54 \\
32 & 0.279 & 0.202 & 0.72$\times$ & 9.01 & 6.52 \\
64 & 0.575 & 0.413 & 0.72$\times$ & 9.12 & 6.56 \\
128 & 1.145 & 0.835 & 0.73$\times$ & 9.02 & 6.57 \\
256 & 2.303 & 1.700 & 0.74$\times$ & 9.03 & 6.67 \\
512 & 4.621 & 3.449 & 0.75$\times$ & 9.04 & 6.75 \\
1024 & 9.272 & 7.128 & 0.77$\times$ & 9.06 & 6.97 \\
2048 & 18.507 & 14.980 & 0.81$\times$ & 9.04 & 7.32 \\
\bottomrule
\end{tabular}
\end{table}

\end{document}